\documentclass[notitlepage,nofootinbib,preprintnumbers,aps,prd, superscriptaddress]{revtex4-1}
\usepackage{amsmath,amssymb,natbib,bm,color}
\usepackage{appendix}
\usepackage{graphicx}

\begin{document}

\title{Reionization in the Light of Dark Stars}
\author{Paolo Gondolo}
\affiliation{Department of Physics and Astronomy, University of Utah, Salt Lake City, UT 84112, USA}
\affiliation{Department of Physics, Tokyo Institute of Technology, Tokyo 152-8551, Japan}
\affiliation{Kavli Institute for the Physics and Mathematics of the Universe, The University of Tokyo, Kashiwa, Chiba 277-8583, Japan}
\author{Pearl Sandick}
\affiliation{Department of Physics and Astronomy, University of Utah, Salt Lake City, UT 84112, USA}
\author{Barmak Shams Es Haghi}
\affiliation{Department of Physics and Astronomy, University of Utah, Salt Lake City, UT 84112, USA}
\author{Eli Visbal}
\affiliation{Department of Physics and Astronomy and Ritter Astrophysical Research Center, University of Toledo, Toledo, OH 43606, USA}

\begin{abstract}
We investigate the effect of Dark Stars (DSs) on the reionization history of the Universe, and the interplay between them and feedback due to Lyman-Werner (LW) radiation in reducing the Cosmic Microwave Background (CMB) optical depth to a value within the $\tau = 0.054 \pm 0.007$ range measured by \textit{Planck}. We use a semi-analytic approach to evaluate reionization histories and CMB optical depths, which includes Population II (Pop II) stars in atomic cooling halos and Pop III stars in minihalos with LW feedback, preceded by a DS phase. We show that while LW feedback by itself can reduce the integrated optical depth to the last scattering surface to $\sim 0.05$ only if the Pop III star formation efficiency is less than $\sim 0.2\%$, the inclusion of a population of DSs can naturally lead to the measured CMB optical depth for much larger Pop III star formation efficiencies $\gtrsim 1\%$.

\end{abstract}

\maketitle

\section{introduction}
\label{sec:introduction}
After recombination of electrons and protons in the Universe into neutral hydrogen at redshift $z\sim 1100$, thermal photons decouple and propagate from the surface of last scattering to form the cosmic microwave background (CMB). The CMB redshifts uninterruptedly until the first ionizing sources, which are generally considered to be the first stars, form within galaxies at redshifts $z\lesssim 30$~\cite{Gnedin:2000uj, Ciardi:1999mx, Bromm:2000nn, Wyithe:2003rr, Schaerer:2001jc, Tumlinson:2002ur, Benson:2005cc}. Scattering of the CMB by the reionized intergalactic medium results in an integrated optical depth $\tau$. Explanations of the \textit{Planck} measured value $\tau=0.054\pm0.007$ (68\% CL)~\cite{Planck:2018vyg} with standard metal-free Population III (Pop III) stars struggle with overproduction of ionizing radiation for high star formation efficiencies. In this study we show that the measured optical depth can be naturally achieved by replacing some or all of the first Pop III stars with Dark Stars (DSs), which are powered by dark matter (DM) annihilation instead of nuclear fusion~\cite{Spolyar:2007qv}. In contrast to more standard Pop III stars, DSs produce a negligible amount of ionizing radiation, and typically form Pop III stars at the end of their lives.

Though the nature of the first stars is, as yet, unknown, any hard ionizing radiation produced by them would have reionized the neutral gas in the intergalactic medium (IGM), a process which is completed by redshift $z\sim 7$~\cite{Fan:2005es, Dawson:2007bz}; afterwards the IGM remains fully-ionized until today. The scattering of CMB photons off the resultant free electrons from reionization modifies the anisotropy power spectrum of the CMB observed today. This information is encoded in the integrated optical depth to the surface of last scattering, $\tau$. 

Pop III stars presumably form from metal-free gas consisting of primordial hydrogen and helium synthesized in the early Universe.  Pop II star formation is thought to take place
in DM minihalos with masses of $\sim 10^6 M_{\odot}$~\cite{Haiman:1996wue, Tegmark:1996yt, Abel:2001pr, Bromm:2001bi}.
Formation of Pop III stars in minihalos demands efficient molecular cooling, because primordial gas does not have the cooling pathways due to transitions atomic/molecular energy levels provided by metals. The cooling process can even become less efficient since formation of Pop III stars is accompanied by emission of Lyman-Werner (LW) photons with energy in the $11.2-13.6\,\textrm{eV}$ range, which can dissociate molecular hydrogen~\cite{Haiman:1996rc, Machacek:2000us, Wise:2007cf, OShea:2007ita, Visbal:2014fta, 2011MNRAS.418..838W}.
Because of the photo-dissociation of molecular hydrogen, eventually only halos with virial temperatures $T_{\textrm{vir}}\gtrsim10^4\,\textrm{K}$ (atomic cooling halos) can form stars by atomic hydrogen cooling. Formation of Pop III stars continues in these halos until they become metal-rich at the onset of Pop I/II star formation.
Pop III are thought to be able to contribute significantly to reionization
in that they have been shown to be more efficient at producing ionizing radiation than metal-enriched stars, i.e., Pop I/II stars~\cite{Tumlinson:1999iu, Schaerer:2001jc, Schaerer:2002yr}.

The small value of the CMB electron scattering optical depth measured by {\it Planck}, $\tau=0.054\pm 0.007$ (68\% CL)~\cite{Planck:2018vyg}, which is consistent with the decreasing trend of the previous measurements by {\it Planck}~\cite{Planck:2015fie} and {\it WMAP}~\cite{2011ApJS..192...18K},
makes early reionization even more challenging than before. A novel idea to delay early star formation and reduce early partial reionization, first put forward in~\cite{Scott:2011ni}, is to consider the affect of annihilation of DM particles, such as weakly-interacting massive particles (WIMPs), into  Standard Model particles at the center of minihalos, as in~\cite{Spolyar:2007qv} and subsequent works.
During star formation, baryons steepen the gravitational potential within the minihalo after cooling and contracting, which draws more DM into its center. This leads to a spike in the DM annihilation rate, followed by injection of a significant amount of energy into the collapsing baryon cloud, which halts or delays star formation. 
DM annihilation at the core of minihalos results in a DS, a partially-collapsed and cool object~\cite{Spolyar:2007qv, Natarajan:2008db}. For a review of DS formation, we refer to~\cite{Freese:2015mta} and references therein.

In this study, we present a new calculation of the effect of DSs on the ionization history and the corresponding integrated optical depth to the last scattering surface.
The effect of the LW background on reducing the CMB optical depth by increasing the minimum critical halo mass required for cooling, which has been investigated in detail in Ref.~\cite{Visbal:2015rpa}, is also considered here.
By exploring the interplay between DSs and LW feedback, we show that while LW feedback can barely reduce the CMB optical depth to meet the constraints set by {\it Planck} data for Pop III star formation efficiency less than $\sim 10^{-4}$, DSs can easily decrease the CMB optical depth to satisfy the data even for Pop III star formation efficiency as high as $\sim 10^{-2}$. 

The outline of this paper is as follows. In Section~\ref{sec:reionization}, we describe our reionization model, which includes LW feedback and a description of the relevant model parameters and their fiducial values. In Section~\ref{sec:addingdarkstars}, after a brief review of DSs and their impact on the reionization history, we modify the reionization model to incorporate them. Finally, in Section~\ref{sec:results} we present and discuss our results, including the effects of DSs on reionization and, consequently, on the integrated optical depth to the last scattering surface, as well as the sensitivity of the optical depth to astrophysical parameters. In this study, a $\Lambda \text{CDM}$ cosmology with following values of the cosmological parameters is assumed: $\Omega_\text{m}h^2=0.14, \Omega_\text{b} h^2=0.022, h=0.67, \sigma_8=0.81$, and $n_\text{s}=0.96$~\cite{Planck:2018vyg}.

\section{reionization model}
\label{sec:reionization}
In this section we review the semi-analytic reionization model presented in~\cite{Visbal:2015rpa} 
with LW feedback but in the absence of DSs.  The inclusion of DSs is described in Section~\ref{sec:addingdarkstars}.

The total ionized filling factor, $Q(z)$, is given by
\begin{equation}
   Q(z)=\rho_\text{b}(z)\int_\infty^zdz'\left[\epsilon_\text{a}\frac{dF_\text{coll,i}}{dz}(z')+(1-Q(z'))\left(\epsilon_\text{a}\frac{dF_\text{coll,a}}{dz}(z')+\epsilon_\text{m}\frac{dF_\text{coll,m}}{dz}(z')\right)\right]V(z',z),
    \label{eq:reionfrac}
\end{equation}
where $\rho_b(z)$ is the mean cosmic baryon density, $V$ is the volume of the ionized region, $\epsilon_\text{\rm \{m,a\}}$ represent the ionizing efficiency, and $F_{\rm coll,\{m,a,i\}}$ are the total fraction of the mass in the Universe collapsed into DM halos at redshift $z'$, where the subscripts $\rm m$, $\rm a$, and $\rm i$ indicate molecular hydrogen cooling halos, atomic hydrogen cooling halos, and halos above the ionized IGM cutoff.  In this model, it is assumed that all molecular hydrogen cooling halos host Pop III stars and all atomic hydrogen cooling halos host Pop II stars.
Specifically, $\epsilon_\text{a}=f_{*,\text{a}}f_\text{esc,a}\eta_\text{ion,a}$ is the ionizing efficiency of atomic cooling halos hosting Pop II stars, and $\epsilon_\text{m}=f_{*,\text{m}}f_\text{esc,m}\eta_\text{ion,m}$ is the ionizing efficiency in minihalos that host Pop III stars. These factors $\epsilon_\text{\rm \{m,a\}}$ count the number of ionizing photons escaping into the IGM per baryon contained in a DM halo and depend on the star formation efficiency (the fraction of baryons in minihalos that form stars), $f_{*,\text{m}}$, the ionizing photon escape fraction, $f_\text{esc,m}$, and the number of ionizing photons produced per baryon included in stars, $\eta_\text{ion,m}$. The fiducial value of these parameters in this paper are as follows: $f_{*,\text{m}}=0.001$, $f_\text{esc,m}=0.5$~\cite{Wise:2014vwa}, $\eta_\text{ion,m}=80000$~\cite{Schaerer:2001jc}, $f_\text{esc,a}=0.15$~~\cite{Wise:2014vwa}, $\eta_\text{ion,a}=4000$~\cite{Samui:2006nk}, and $f_{*,\text{a}}=f_{*,\text{a}}(z)$ is taken from Ref.~\cite{Visbal:2015rpa}.

In this model, the number of DM halos per unit comoving volume of the Universe is evaluated analytically with the Sheth-Tormen mass function~\cite{Sheth:1999mn}.
In the regions of the IGM that have already been ionized, star formation below a halo mass scale, $M_\text{i}(z)=1.5\times10^8(\frac{1+z}{11})^{-1.5} M_{\odot}$~\cite{Dijkstra:2003vg}, is prevented by the increased Jeans mass of the heated gas. 
The masses of atomic cooling halos that host Pop II stars are assumed to be larger than $M_\text{a}(z)=5.4\times10^7(\frac{1+z}{11})^{-1.5} M_{\odot}$~\cite{Fernandez:2014wia}.
The minimum mass of minihalos hosting Pop III stars, which is sensitive to the LW background (flux), $J_\text{LW}(z)$, and the baryon-DM streaming velocity, $v_\text{bc}$, will be discussed later.

The total fraction of the mass in the Universe that is collapsed into
DM halos above the ionized IGM cutoff, $F_\text{coll,i}(z)$, atomic cooling halos, $F_\text{coll,a}(z)$, and minihalos, $F_\text{coll,m}(z)$, are given by
\begin{eqnarray}
 \nonumber F_\text{coll,i}(z)&=&\frac{1}{\Omega_\text{m}\rho_\text{c}}\int_{M_\text{i}}^\infty dM M\frac{dn}{dM}(z),\\ 
 \nonumber F_\text{coll,a}(z)&=&\frac{1}{\Omega_\text{m}\rho_\text{c}}\int_{M_\text{a}}^{M_\text{i}} dM M\frac{dn}{dM}(z),\\
   F_\text{coll,m}(z)&=&\frac{1}{\Omega_\text{m}\rho_\text{c}}\int_{M_\text{m}}^{M_\text{a}} dM M\frac{dn}{dM}(z),
   \label{eq:Fcoll}
\end{eqnarray}
respectively, where $\rho_\text{c}$ is the critical energy density and $dn(z)/dM$ is the Sheth-Tormen mass function.

The factor $1-Q(z)$ is included to make sure that new ionizing sources appear in minihalos and atomic cooling halos only in regions that have not yet been ionized~\cite{Haiman:2003ea}. $V(z',z)$ represents the expanding ionized region into the IGM where the corresponding halo has formed at $z'$ ($z\lesssim z'$). The evolution of the ionization front, $R_\text{i}=\left[3V/(4\pi)\right]^{1/3}$, is governed by
\begin{equation}
    \frac{dR^3_\text{i}}{dt}=3H(z)R^3_\text{i}+\frac{3\dot{N}_\gamma}{4\pi\langle n_H\rangle}-C(z)\langle n_H\rangle\alpha_\text{B}R^3_\text{i},
\end{equation}
where $H(z)$ is the Hubble expansion rate, $\langle n_H\rangle$ is the mean hydrogen density in Universe, and $C(z)\equiv \langle n^2_\textrm{ HII}\rangle/\langle n_\textrm {HII}\rangle^2=2\left(\frac{1+z}{7}\right)^{-2}+1$ is the clumping factor of the ionized IGM~\cite{Bauer:2015tta}. $\alpha_\text{B}=2.6 \times 10^{-13}\,\text{cm}^3\text{s}^{-1}$ is the case B (optically-thick) recombination coefficient of hydrogen at $T=10^4 \,\text{K}$. The rate of ionizing photon emission, for each solar mass of star forming gas with ionizing efficiency normalised to one, $\dot{N}_\gamma$, is given by 
\begin{equation}
    \dot{N}_\gamma(t)=\dot{N}_0\left[\theta(t_{6.5}-t)+(t/t_{6.5})^{-4.5}\theta(t-t_{6.5})\right],
    \label{eq:Ngamma}
\end{equation}
where $\dot{N}_0=9.25\times 10^{42}\,\text{s}^{-1}M_{\odot}^{-1}$, $\theta(t)$ is the unit step function, $t_{6.5}=10^{6.5}\text{yr}$, and $t$ is measured after starburst. This rate of ionizing photon emission results in one ionizing photon per baryon incorporated into stars over the  lifetime of the stellar population.

The minimum mass of minihalos hosting Pop III star formation, $M_\text{m}$, is affected by LW radiation and the baryon-DM streaming velocity.  Increased LW radiation leads to an increased fraction of dissociated molecular hydrogen, which increases $M_\text{m}$.  Similarly, an increased streaming velocity delays the inflow of gas into halos~\cite{2010PhRvD..82h3520T}, which also increases $M_\text{m}$~\cite{10.1111/j.1365-2966.2012.21212.x, 10.1111/j.1365-2966.2012.20605.x, doi:10.1142/S0218271814300171, 10.1093/mnras/stz013}.
These two effects are included in Eq~(\ref{eq:reionfrac}) via $M_\text{m}$ in Eq.~(\ref{eq:Fcoll}). In this study, we consider two functional forms for the dependence of $M_\text{m}$ on redshift,  both based on hydrodynamical cosmological simulations. In the first form, the effect of the baryon-DM streaming velocity is ignored, and $M_\text{m}$ depends only on the density of LW radiation, $J_\text{LW}$~\cite{2013MNRAS.432.2909F}:
\begin{equation}
    M_\text{m}(J_\text{LW},z)=2.5\times10^5\left(\frac{1+z}{26}\right)^{-1.5}\left[1+6.96(4\pi J_\text{LW}(z))^{0.47}\right]M_{\odot}.
\end{equation}
In the second form, the baryon-DM streaming velocity $v_\text{bc}$ is included, and
the following fit formula is used for $M_\text{m}$~\cite{Kulkarni:2020ovu}:
\begin{equation}
M_\text{m}(J_\text{LW},v_\text{bc},z)=1.96\times 10^5(1+J_\text{LW}(z))^{0.8}\left(1+\frac{v_\text{bc}}{30}\right)^{1.83}\left(1+ \frac{J_\text{LW}(z)v_\text{bc}}{3}\right)^{-0.06}\left(\frac{1+z}{21}\right)^{\alpha(J_\text{LW},v_\text{bc})}M_{\odot}.
    \label{eq:McritJV}
\end{equation}
Here
\begin{equation}
    \alpha(J_\text{LW},v_\text{bc})={-1.64(1+J_\text{LW}(z))^{0.36}\left(1+\frac{v_\text{bc}}{30}\right)^{-0.62}\left(1+ \frac{J_\text{LW}(z)v_\text{bc}}{3}\right)^{0.13}},
    \label{alpha}
\end{equation}
$J_\text{LW}$ is measured in units of $10^{-21}\,\text{erg}\,\text{s}^{-1}\text{cm}^{-2}\text{Hz}^{-1}\text{Sr}^{-1}$, and $v_\text{bc}$ is measured in km/s. The fiducial value of the baryon-DM streaming velocity is assumed to be $30\,\text{km/s}$~\cite{Kulkarni:2020ovu}.
For this fit formula, simulations have been updated such that they include the effect of molecular hydrogen self-shielding, which acts to lower the minimum halo mass for Pop III star formation (the opposite of the effect of the streaming velocity)~\cite{Kulkarni:2020ovu}.

Provided that the IGM is almost transparent to LW photons until they are redshifted into a Lyman series line and absorbed, or equivalently by assuming that at redshift $z$ all LW photons emitted from sources at $1.015 z$ are observable (since an LW photon can redshift by 1.5\% before reaching a Lyman series), the intensity of the LW background can be evaluated by
\begin{equation}
    J_\text{LW}(z)=\frac{c(1+z)^3}{4\pi}\int_{1.015 z}^zdz'\frac{dt_\text{H}}{dz'}\left(\frac{\text{SFRD}_\text{a}(z')}{m_\text{p}}\eta_\text{LW,a}+\frac{\text{SFRD}_\text{m}(z')}{m_\text{p}}\eta_\text{LW,m}\right)E_\text{LW}\Delta\nu^{-1}_\text{LW},
\end{equation}
where $c$ is the speed of light, $t_\text{H}$ is the Hubble time, and $\text{SFRD}_\text{a,m}(z)$  are the star formation rate densities for atomic (a) and molecular (m) cooling halos, given by
\begin{eqnarray}
\nonumber\text{SFRD}_\text{a}(z)&=&\rho_\text{b}f_{*,\text{a}}\frac{dF_\text{coll,a}}{dt}\left(1-Q(z)\right)+\rho_\text{b}f_{*,\text{a}}\frac{dF_\text{coll,i}}{dt},\\
\text{SFRD}_\text{m}(z)&=&\rho_\text{b}f_{*,\text{m}}\frac{dF_\text{coll,m}}{dt}\left(1-Q(z)\right).
\end{eqnarray}
Here, $m_\text{p}$ is the proton mass, $\eta_\text{LW}$ counts the number of LW photons per baryon produced by stars, $E_\text{LW}=1.9\times 10^{-11}\,\text{erg}$, and $\Delta\nu_\text{LW}=5.8\times 10^{14} \, \text{Hz}$.

After solving Eq.~(\ref{eq:reionfrac}) iteratively for $Q(z)$ and $J_\text{LW}(z)$, the ionized filling factor can be used to evaluate the optical depth as
\begin{equation}
    \tau(z)=\int_0^z dz'\frac{c(1+z')^2}{H(z')}Q(z')\sigma_{\textrm{T}}\langle n_H\rangle\left(1+\eta_{\textrm{He}}(z')\frac{Y}{4X}\right),
\end{equation}
where $\sigma_{\textrm{T}}$ is the Thompson scattering cross section, $Y=0.24$ and $X=0.76$ are the mass fractions of helium and hydrogen respectively, and $\eta_{\textrm{He}}(z)=\theta(z-3)+2\theta(3-z)$, provided that helium and hydrogen are singly ionized at the same time while helium is doubly ionized at $z = 3$.
\section{adding Dark Stars}
\label{sec:addingdarkstars}
The impact of DSs on the reionization history has been studied in Ref.~\cite{Scott:2011ni}, which we briefly review here. 
The role of DM in the formation of Pop~III stars is not limited to providing potential wells for baryonic collapse. It has been shown that DM annihilation into Standard Model particles may grow drastically when baryons contract within minihalos during star formation, and may result in the formation of a new phase of stars, called DSs, powered by the annihilation of DM inside them rather than nuclear fusion~\cite{Spolyar:2007qv}. DSs have low surface temperatures and do not emit any relevant amount of ionising radiation. Eventually, when the DM annihilation runs out, the DS phase ends with the star becoming either a Pop III star or, if very massive, exploding~\cite{Spolyar:2007qv,Natarajan:2008db}. Formation of a DS demands efficient accumulation of DM at the center of the protostar. Gravitational contraction of baryonic gas during the collapse steepens the gravitational potential in the core of the halo, which drags more DM into the center of the cloud. The other way for DM to be captured into the core of stars is by losing kinetic energy through scattering off nucleons in the star. Therefore the evolution and lifetime of DSs depends on the rate of DM accumulation at the core of protostars. 
DM provided by gravitational contraction runs out in $\sim 0.4\,\textrm{Myr}$ while DM capturing by DSs can continue and keep them alive as long as they lie within a region with high enough density of DM~\cite{2009ApJ...705.1031S}. 

For simplicity, we assume that DM annihilation contributes substantially into the energy budget of the star, and that the capture rate is sufficiently large to keep the star cool and make its contribution to reionization almost zero~\cite{Spolyar:2007qv,Scott:2011ni}.
DSs are described by two parameters of interest in this study: the DS mass fraction, $f_{\textrm{DS}}$, that describes the fraction of the baryonic mass that initially goes into DSs rather than Pop III stars, and the lifetime of DSs, $t_{\textrm{DS}}$.

To model the reionization process from star formation in DM halos, we modify the model presented in~\cite{Visbal:2015rpa} to include the effect of DSs on delaying formation of Pop III stars. 
The impact of DSs on reionization is simplified by assuming that DSs, which contain a fraction $f_\text{DS}$ of the baronic mass, halt reionization by delaying star formation in minihalos and atomic cooling halos during their lifetime ($t'\lesssim t\lesssim t'_{\text{DS}} \equiv t'+t_\text{DS})$ and contribute nothing to reionization, but after they run out of DM ($t\gtrsim t'_\text{DS})$, they die and are replaced with either Pop III or Pop II stars. This is achieved by modifying Eq.~\eqref{eq:reionfrac} to

\begin{equation}
    Q(z)=\rho_\text{b}(z)\int_\infty^zdz'\left[\epsilon_\text{a}\frac{dF_\text{coll,i}}{dz}(z')+(1-Q(z'))\left(\epsilon_\text{a}\frac{dF_\text{coll,a}}{dz}(z')+\epsilon_\text{m}\frac{dF_\text{coll,m}}{dz}(z')\right)\phi(z',z'_\text{DS},f_\text{DS})\right]V(z',z),
    \label{eq:reionfracwithDS}
\end{equation}
where
\begin{equation}
\phi(z',z'_\text{DS},f_\text{DS})=(1-f_\text{DS})\theta(z'-z'_\text{DS})+\theta(z'_\text{DS}-z'),
\end{equation}
and $z'_{\text{DS}}$ is the redshift at time $t'_{\text{DS}}$.

\section{results}
\label{sec:results}
In this section, we present the results of the reionization model modified by adding DSs.
In Subsection~\ref{sec:DSeffect}, we show how the total ionized filling factor and subsequently the optical depth of the fiducial reionization model change in the presence of DSs. In Subsection~\ref{sec:Astroeffect}, we explore the effects of varying the astrophysical parameters on the reionization model for a benchmark DS example.

\subsection{Effects of DSs}
\label{sec:DSeffect}
\begin{figure}[t]
  \centering
  \includegraphics[width=0.47\textwidth]{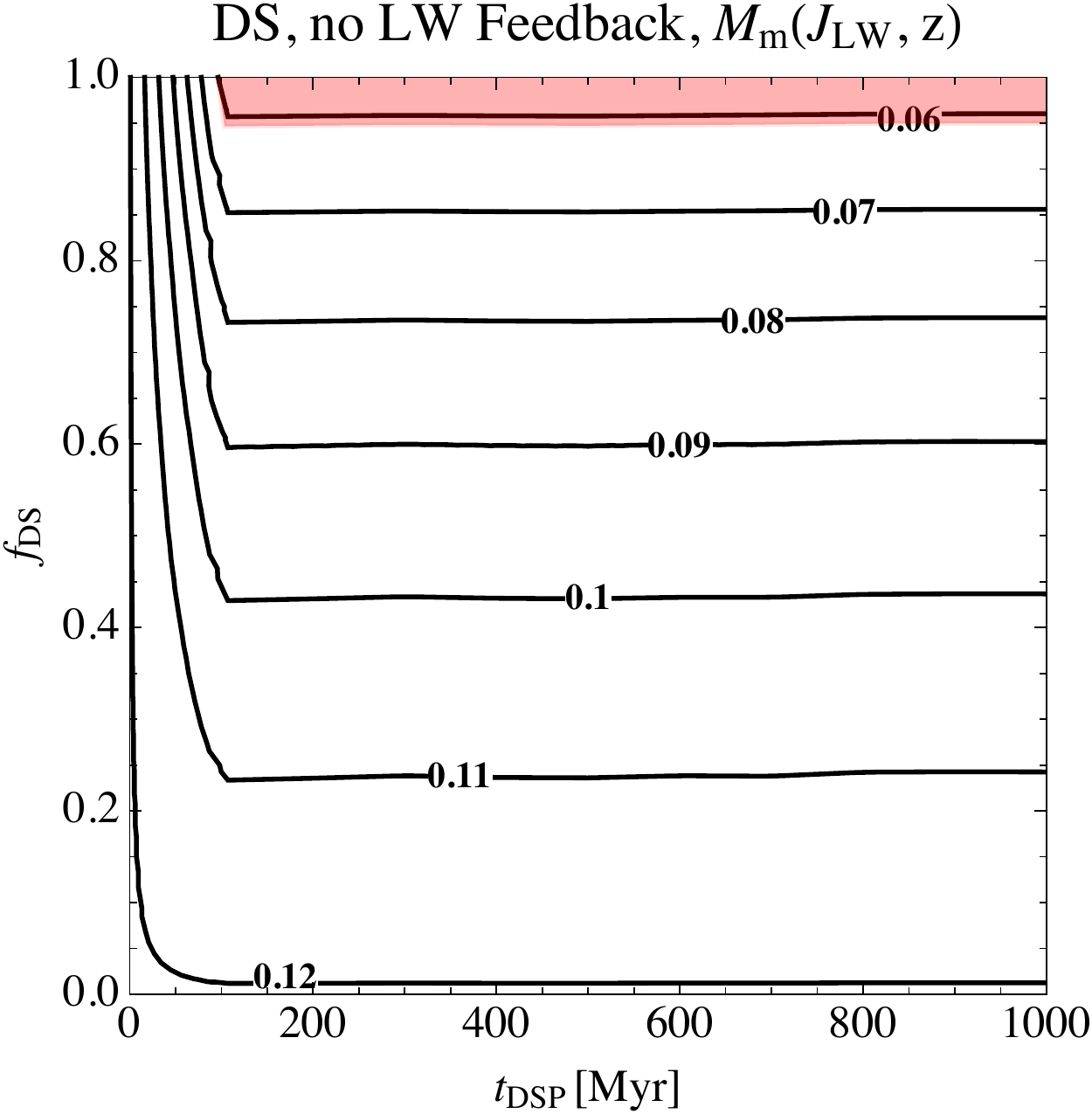}\hspace{3mm}
  \includegraphics[width=0.47\textwidth]{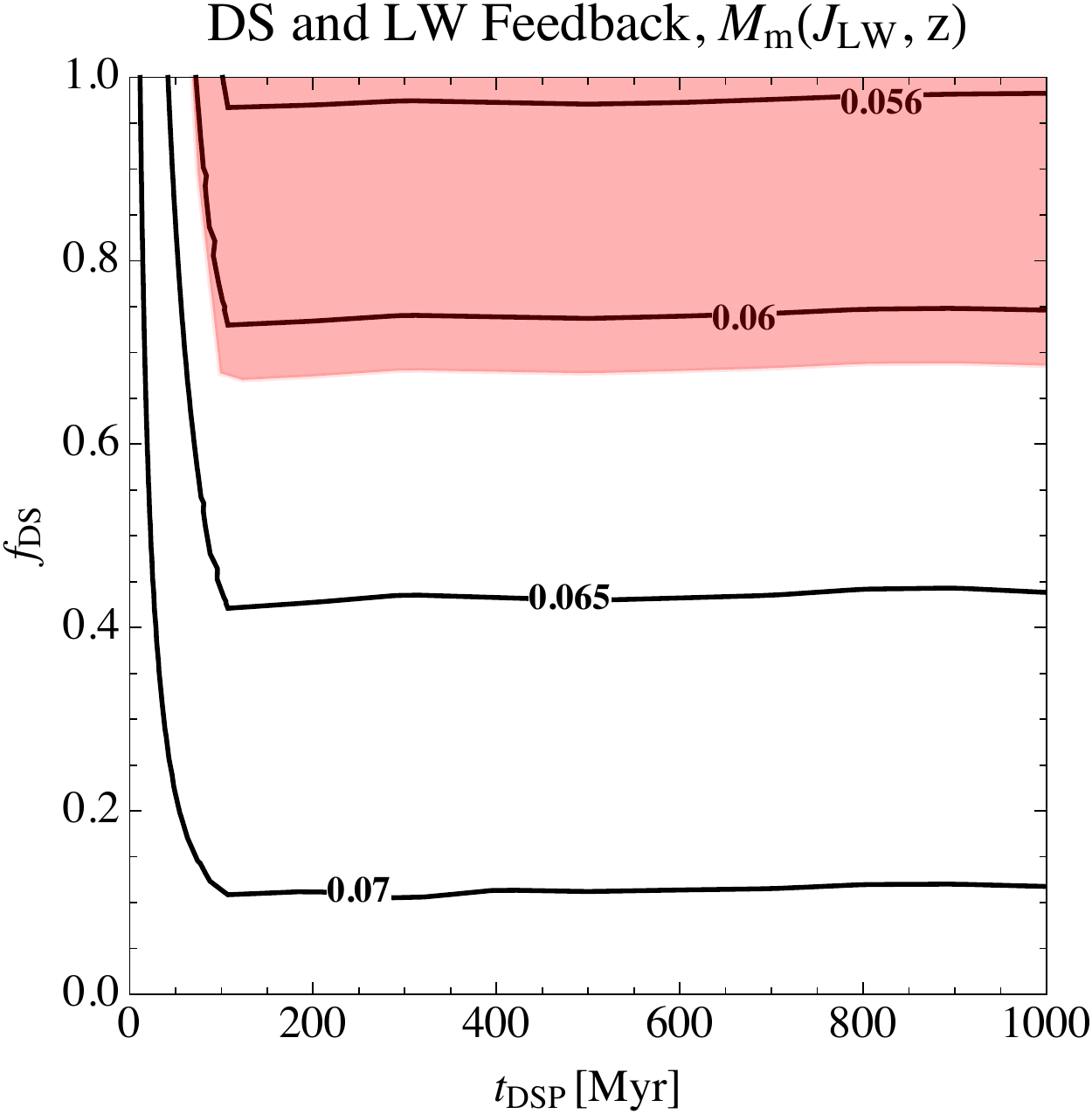}\\\vspace{4mm}
  \includegraphics[width=0.47\textwidth]{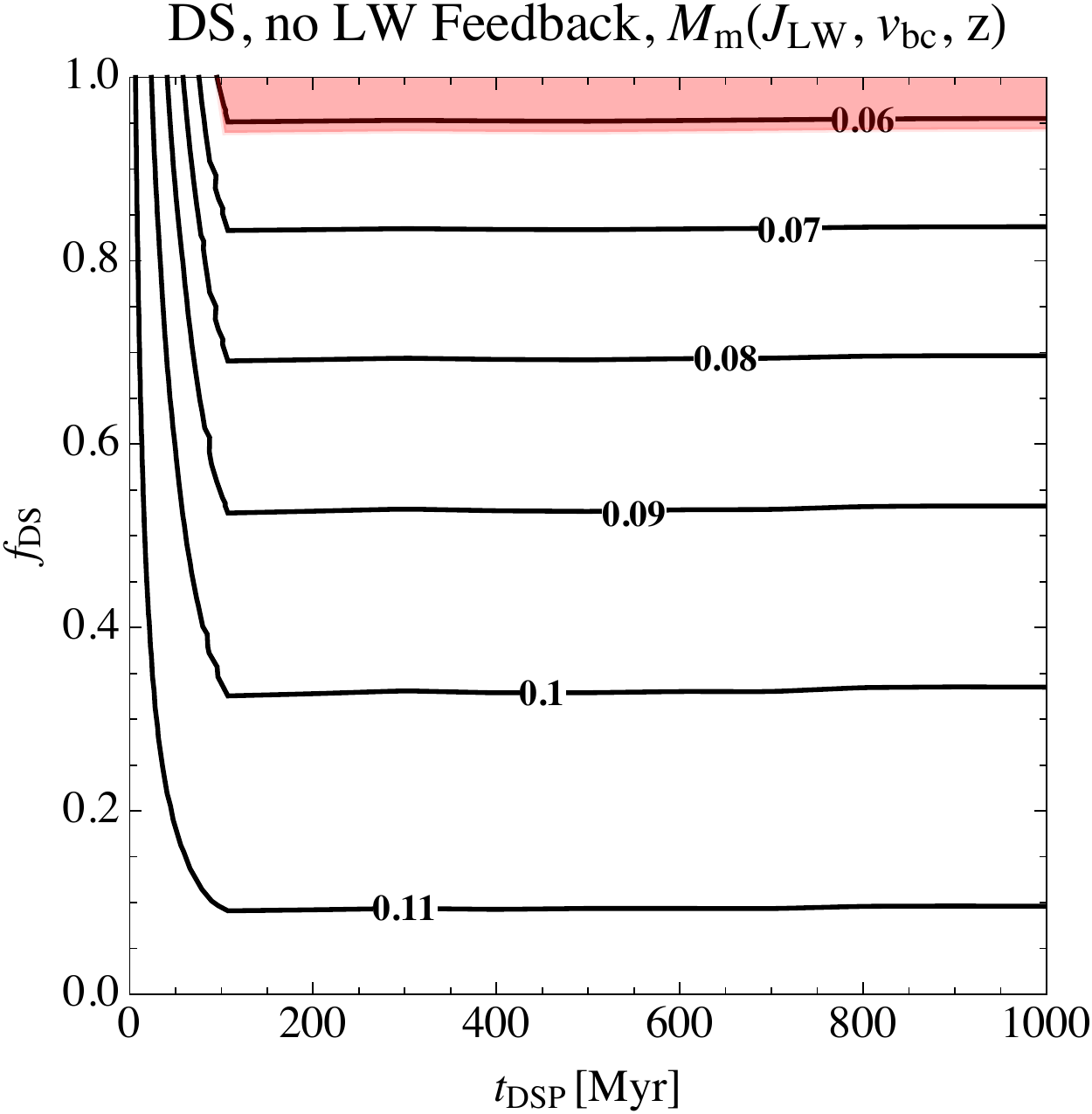}\hspace{3mm}
  \includegraphics[width=0.47\textwidth]{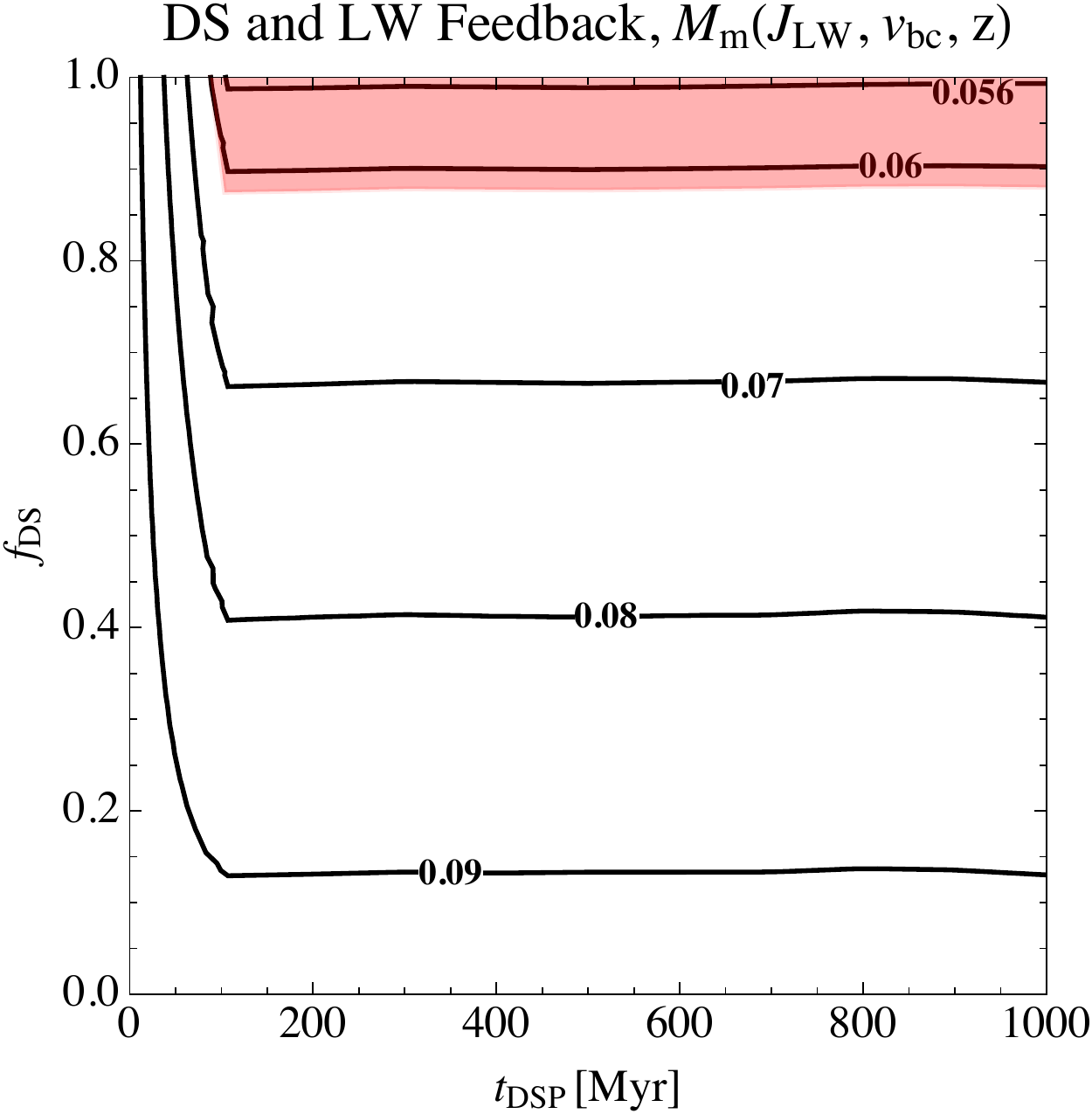}
  \caption{Contours of the integrated optical depth to
the last scattering surface as a function of DSs mass fraction, $f_\text{DS}$, and the lifetime of DSs, $t_\text{DS}$. Top (bottom) panel corresponds to the minimum mass of minihalos,  $M_{\textrm{m}}$, in which the effect of the baryon-DM streaming velocity, is ignored (included). Left (right) panels show the optical depth in presence of DSs when LW feedback is ignored (included). The astrophysical parameters assume their fiducial values. The red shaded regions display $1\sigma$ regions based on the integrated optical depth to the last scattering surface observed by {\it Planck} ($\tau=0.054\pm 0.007$)~\cite{Planck:2018vyg}. }
  \label{fig:tauDS}
\end{figure}

In Fig.~\ref{fig:tauDS}, we display contours of the integrated optical depth to
the last scattering surface as a continuous function of $f_\text{DS}$ and $t_\text{DS}$.
The astrophysical parameters assume their fiducial values listed at the beginning of Section~\ref{sec:reionization}.
The top (bottom) panel corresponds to the representation of $M_{\textrm{m}}$ in which the effect of the baryon-DM streaming velocity is ignored (included). 
The left (right) panels in Fig.~\ref{fig:tauDS} show the optical depth in the presence of DSs when LW feedback is ignored (included).
The red shaded regions in Fig.~\ref{fig:tauDS} (and in the rest of the figures in this paper) display $1\sigma$ regions based on the integrated optical depth to the last scattering surface observed by {\it Planck}, i.e., $\tau=0.054\pm 0.007$ ~\cite{Planck:2018vyg}.

From the left panels of Fig.~\ref{fig:tauDS}, we can see that the two representations of $M_{\textrm{m}}$, in the lack of LW feedback and in the presence of DSs, lead to almost the same result. Namely, DSs with a lifetime in the range $100\,\textrm{Myr}\lesssim t_{\textrm{DS}}\lesssim1000\,\textrm{Myr}$ and with a mass fraction in the range $0.95\lesssim f_{\textrm{DS}}\lesssim 1$, give rise to the optical depth consistent with limits from {\it Planck}. 

The right panels of Fig.~\ref{fig:tauDS}, on the other hand, indicate further suppression of the optical depth after adding the LW feedback, and also show that the effect of LW feedback is stronger for the $M_{\textrm{m}}$ in which the baryon-DM streaming velocity is ignored, i.e., $M_\text{m}(J_\text{LW},z)$, than when it includes the baryon-DM streaming velocity, i.e., $M_\text{m}(J_\text{LW},v_\text{bc},z)$. After including LW feedback, for the choices $M_\text{m}(J_\text{LW},z)$ and $M_\text{m}(J_\text{LW},v_\text{bc},z)$, respectively, DSs  with a lifetime in the range $100\,\textrm{Myr}\lesssim t_{\textrm{DS}}\lesssim1000\,\textrm{Myr}$ and with a mass fraction in the range $0.7\lesssim f_{\textrm{DS}}\lesssim 1$ ($0.9\lesssim f_{\textrm{DS}}\lesssim 1$, resp.) lead to an optical depth consistent with limits from {\it Planck}.

\begin{figure}[t]
  \centering
  \includegraphics[width=0.47\textwidth]{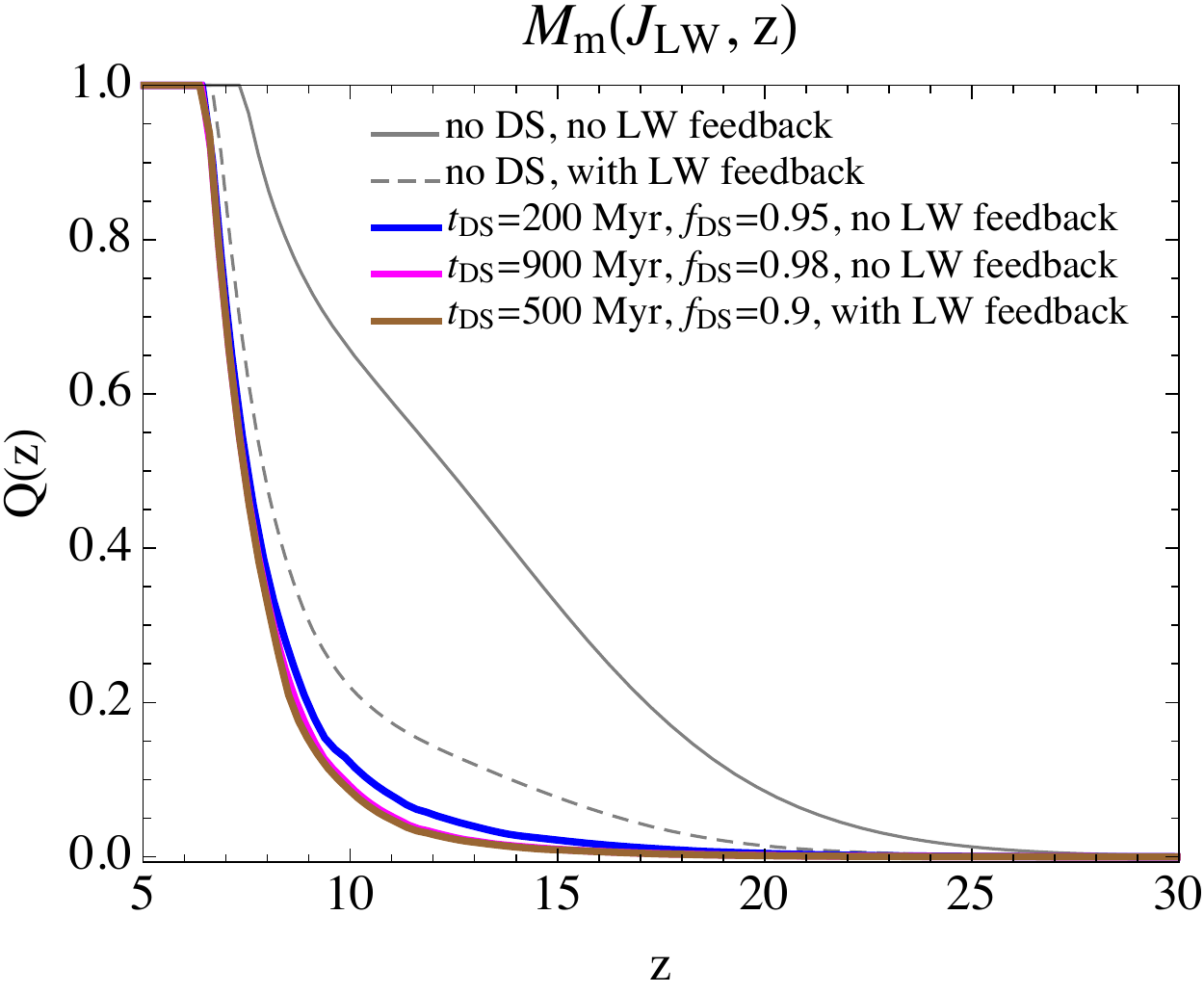}\hspace{3mm}
  \includegraphics[width=0.48\textwidth]{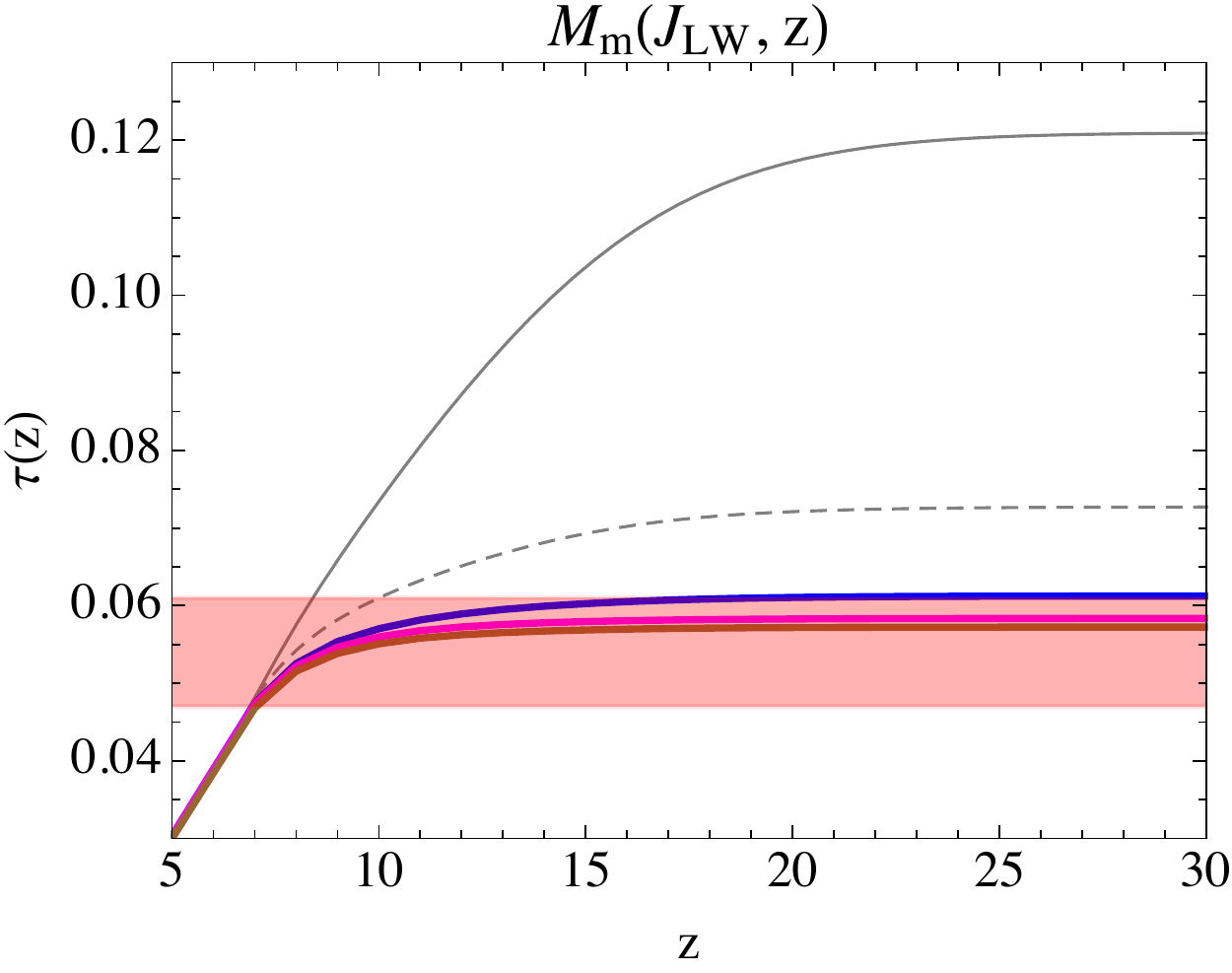}\\\vspace{4mm}
  \includegraphics[width=0.47\textwidth]{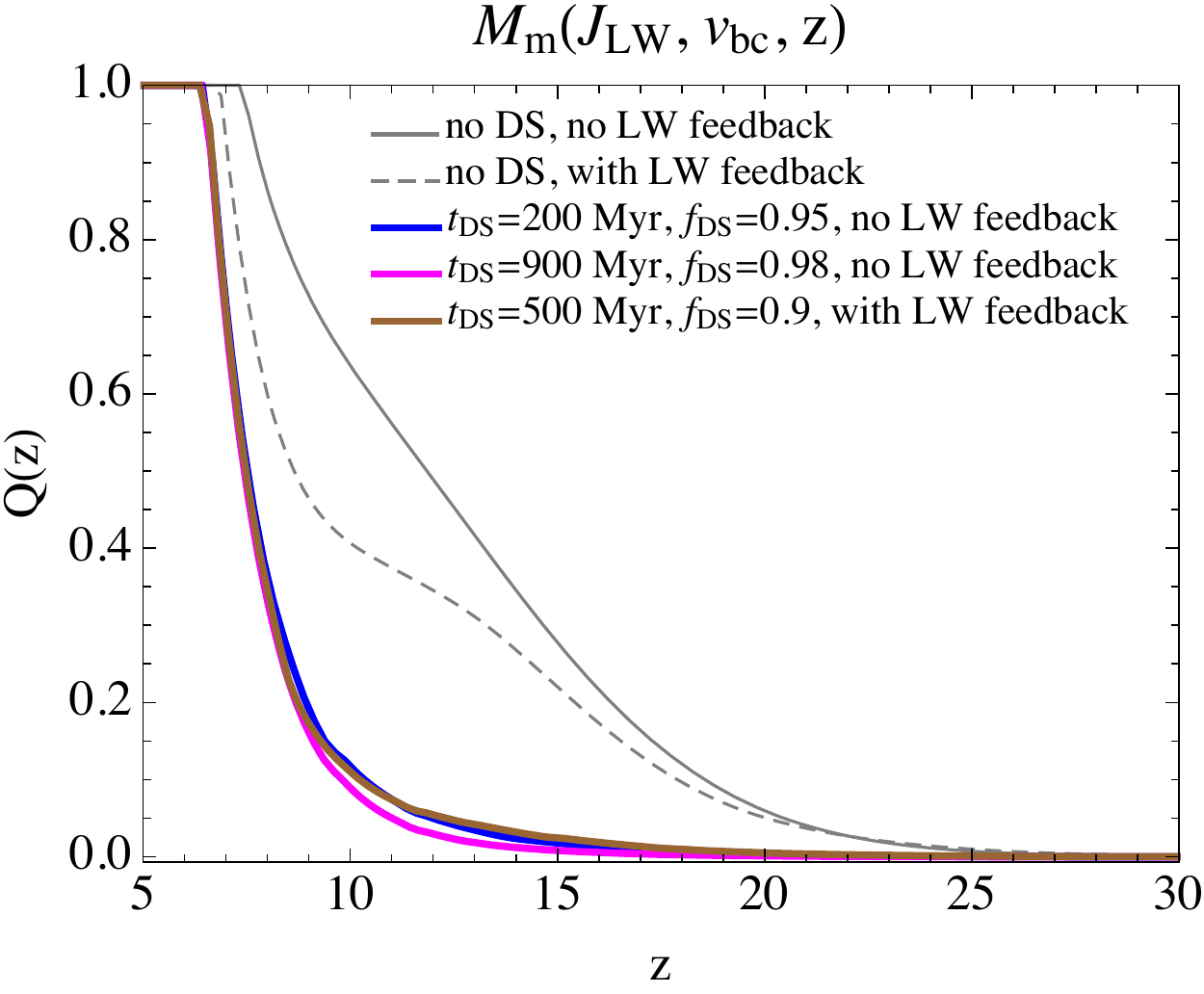}\hspace{3mm}
  \includegraphics[width=0.48\textwidth]{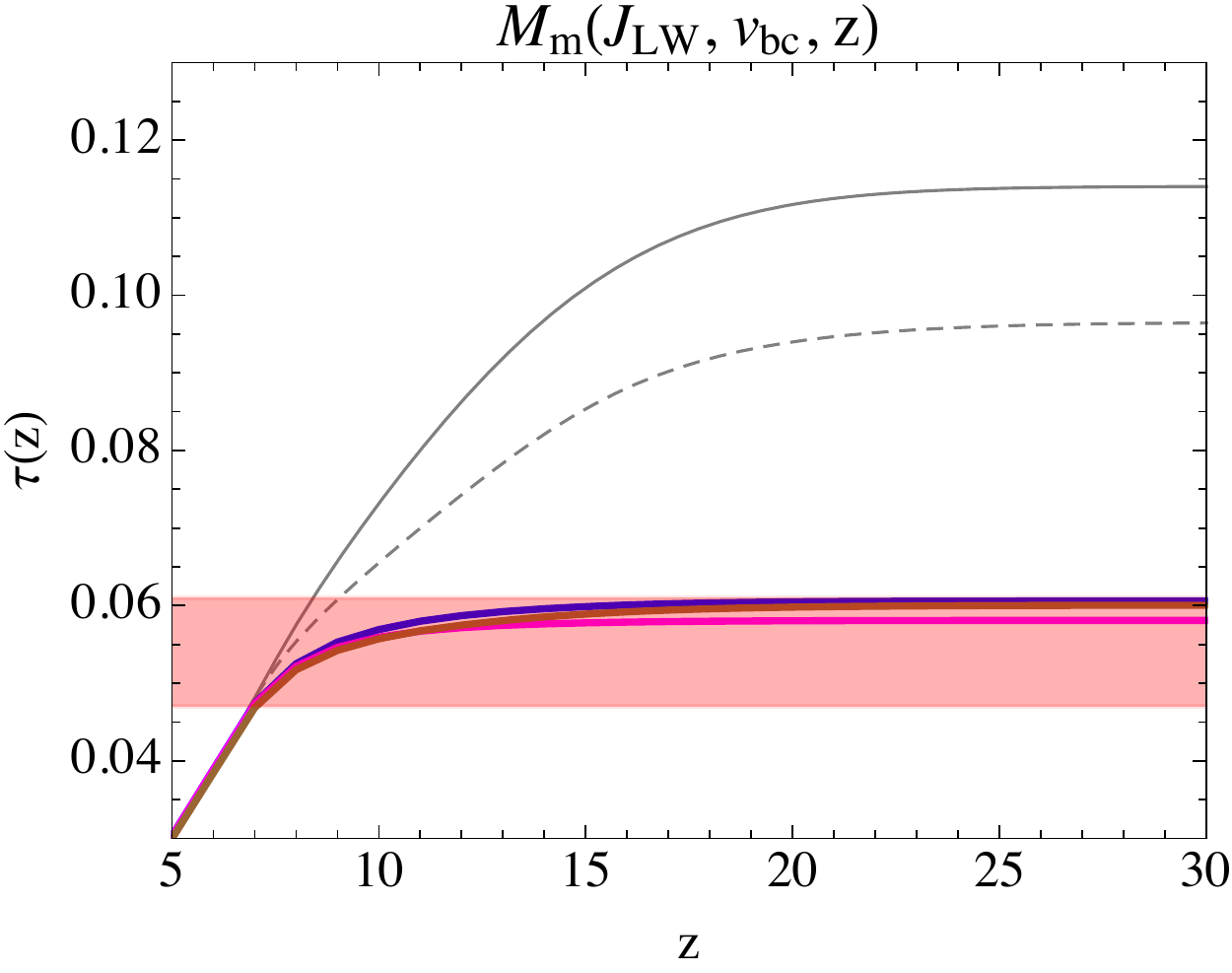}
  \caption{The total ionized filling factor (left panels) and corresponding optical depth from the present day to redshift $z$ (right panels). Top (bottom) panel corresponds to the representation of $M_{\textrm{m}}$ in which the effect of the baryon-DM streaming velocity is ignored (included). Gray solid (dashed) curves corresponds to a reionization history without DSs and in the absence (presence) of LW feedback, blue and magenta curves display reionization histories without LW feedback including DSs with $t_{\text{DS}}=200\,\text{Myr}, f_{\text{DS}}=0.95$ and $t_{\text{DS}}=900\,\text{Myr}, f_{\text{DS}}=0.98$ respectively. The brown curves show reionization in the presence of DSs with $t_{\text{DS}}=500\,\text{Myr}, f_{\text{DS}}=0.9$ by including LW feedback. The red bands in the right panels display $1\sigma$ regions based on the integrated optical depth to the last scattering surface observed by {\it Planck}.}
  \label{fig:Qfunc}
\end{figure}

To elaborate on these results, in Fig.~\ref{fig:Qfunc} we display the total ionized filling factor (left panels) and corresponding optical depth from the present day to redshift $z$ (right panels) for some benchmark DSs.
The top (bottom) panel corresponds to the representation of $M_{\textrm{m}}$ in which the effect of the baryon-DM streaming velocity is ignored (included). DS parameters are chosen such that the resultant optical depth lies within the $1\sigma$ {\it Planck} region. 

The gray solid curve corresponds to a reionization history without DSs and LW feedback. The gray dashed curve depicts the effect of LW feedback in the absence of DSs. The blue and magenta curves correspond to reionization histories which involve DSs with $t_{\text{DS}}=200\,\text{Myr}, f_{\text{DS}}=0.95$ and $t_{\text{DS}}=900\,\text{Myr}, f_{\text{DS}}=0.98$ respectively, without including LW feedback.
The brown curve shows the total ionized filling factor for DSs with $t_{\text{DS}}=500\,\text{Myr}, f_{\text{DS}}=0.9$ in the presence of LW feedback, where these two factors together suppress the optical depth to the acceptable level.
Fig.~\ref{fig:Qfunc} displays clearly the dominance of DSs over LW feedback in decreasing the optical depth.

\subsection{Effects of Astrophysical Parameters}
\label{sec:Astroeffect}
\begin{figure}[t]
  \centering
    \includegraphics[width=0.47\textwidth]{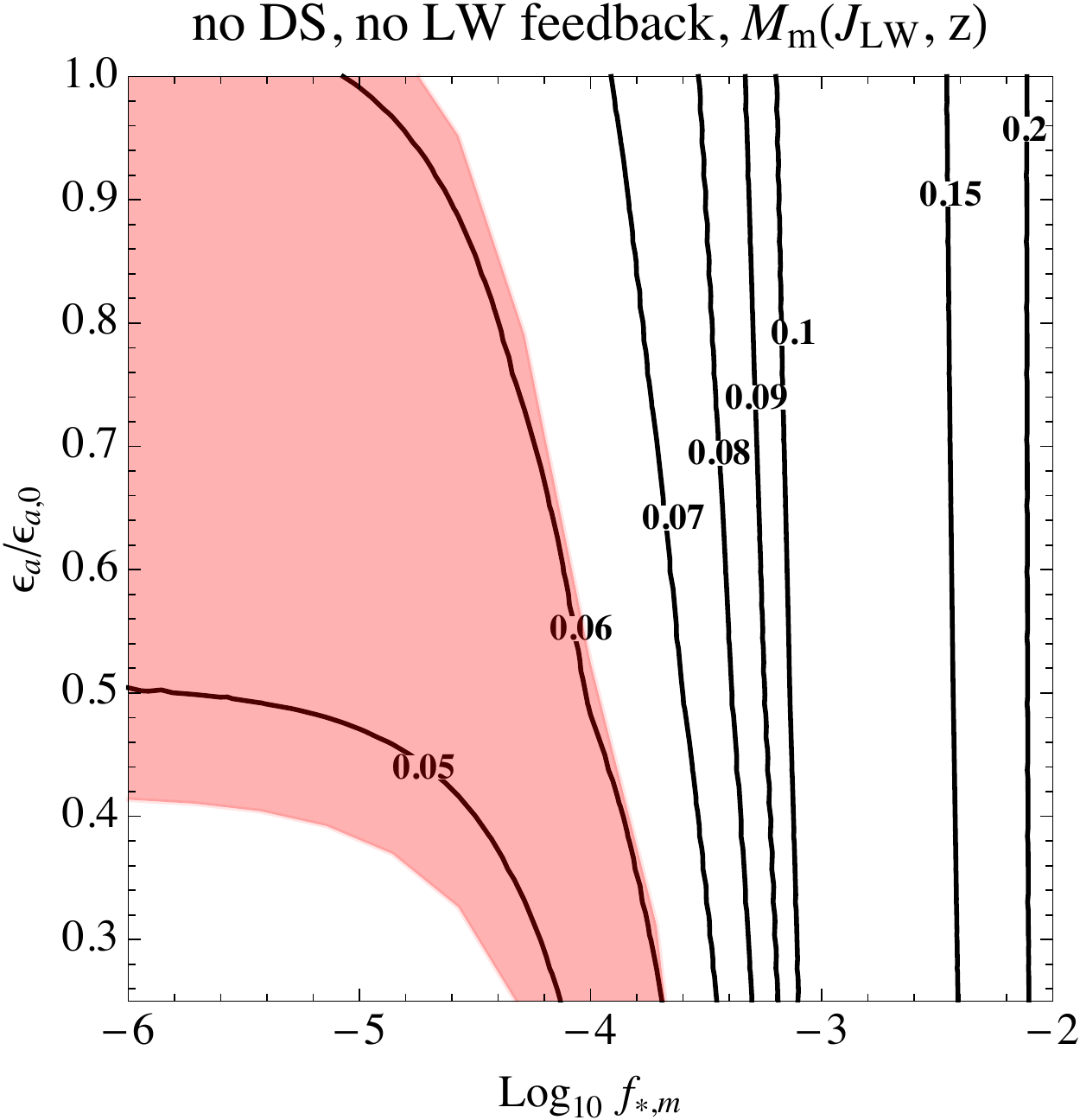}\hspace{3mm}
  \includegraphics[width=0.47\textwidth]{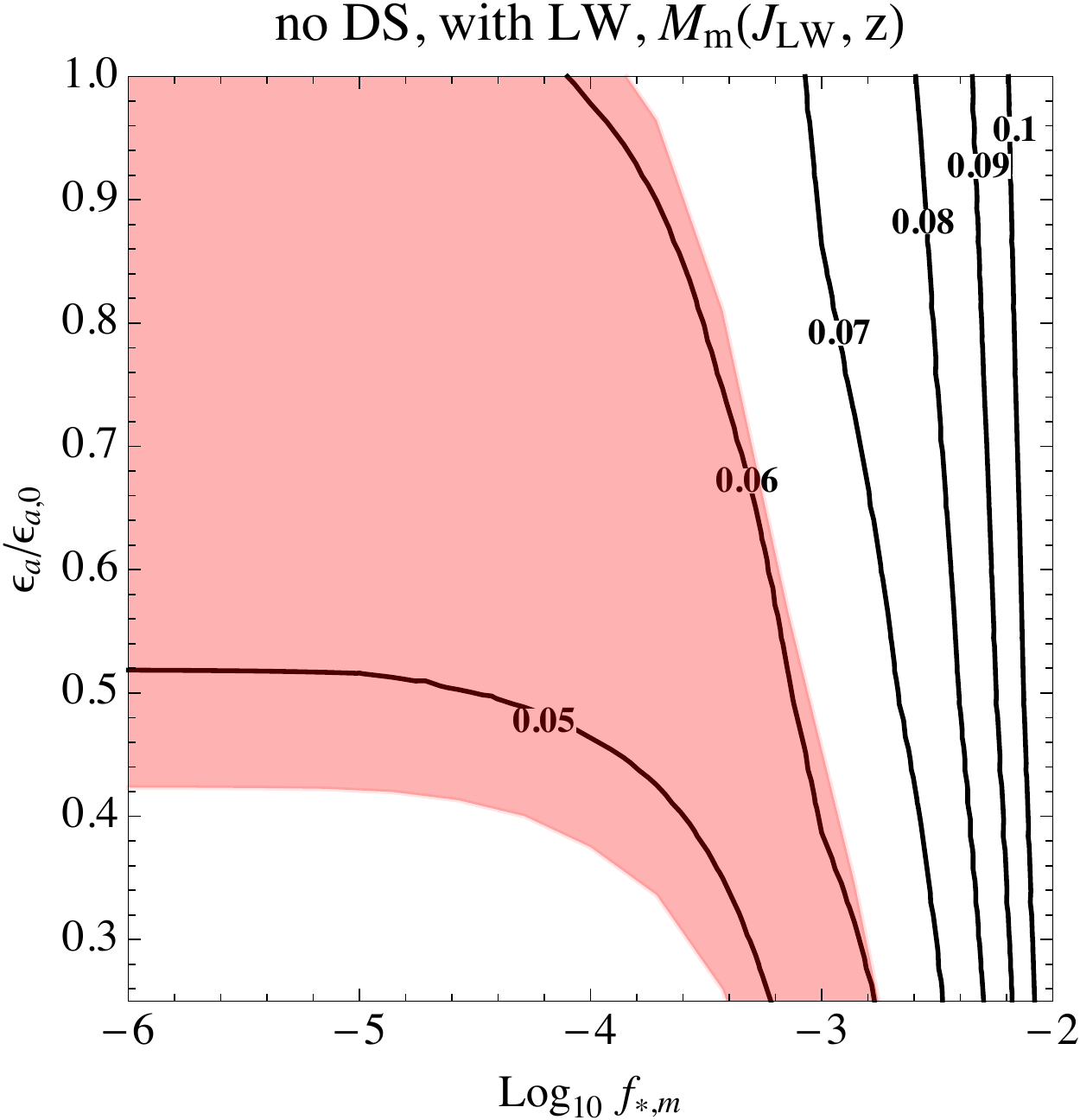}\\
  \caption{contours of the integrated optical depth to
the last scattering surface as a function of $f_{*,\text{m}}$ and $\epsilon_{\text{a}}$, when $M_\text{m}=M_\text{m}(J_\text{LW},z)$.
Left (right) panel show the optical depth without DSs when LW feedback is ignored (included). The red shaded regions display $1\sigma$ regions based on the integrated optical depth to the last scattering surface observed by {\it Planck}.}
  \label{fig:tauAstro1}
\end{figure}

In Fig.~\ref{fig:tauAstro1}, we show contours of the integrated optical depth to
the last scattering surface as a continuous function of $f_{*,\text{m}}$ and $\epsilon_{\text{a}}$, for $ M_\text{m}=M_\text{m}(J_\text{LW},z)$. Dependence on $\epsilon_{\text{a}}$ is represented by the ratio, $\epsilon_{\text{a}}/\epsilon_{{\text{a},0}}\,$, which captures the reduction compared to the fiducial value $\epsilon_{{\text{a},0}}$.
The left (right) panel displays the result in the absence of DSs and without (with) LW feedback.
For $f_{*,\text{m}}\lesssim 2\times10^{-4}$, by selecting a small enough $\epsilon_{\text{a}}$, the resultant integrated optical depth could be consistent with {\it Planck} data without DSs or LW feedback. Including LW feedback makes it possible to increase the value of $f_{*,\text{m}}$ up to $2\times10^{-3}$. It can also be concluded from Fig.~\ref{fig:tauAstro1} that for small values of $f_{*,\text{m}}$, the optical depth depends on both $f_{*,\text{m}}$ and $\epsilon_{\text{a}}$, while larger values of $f_{*,\text{m}}$ dominate over $\epsilon_{\text{a}}$ such that the integrated optical depth is almost independent of $\epsilon_{\text{a}}$.

\begin{figure}[t]
  \centering
  
  \includegraphics[width=0.47\textwidth]{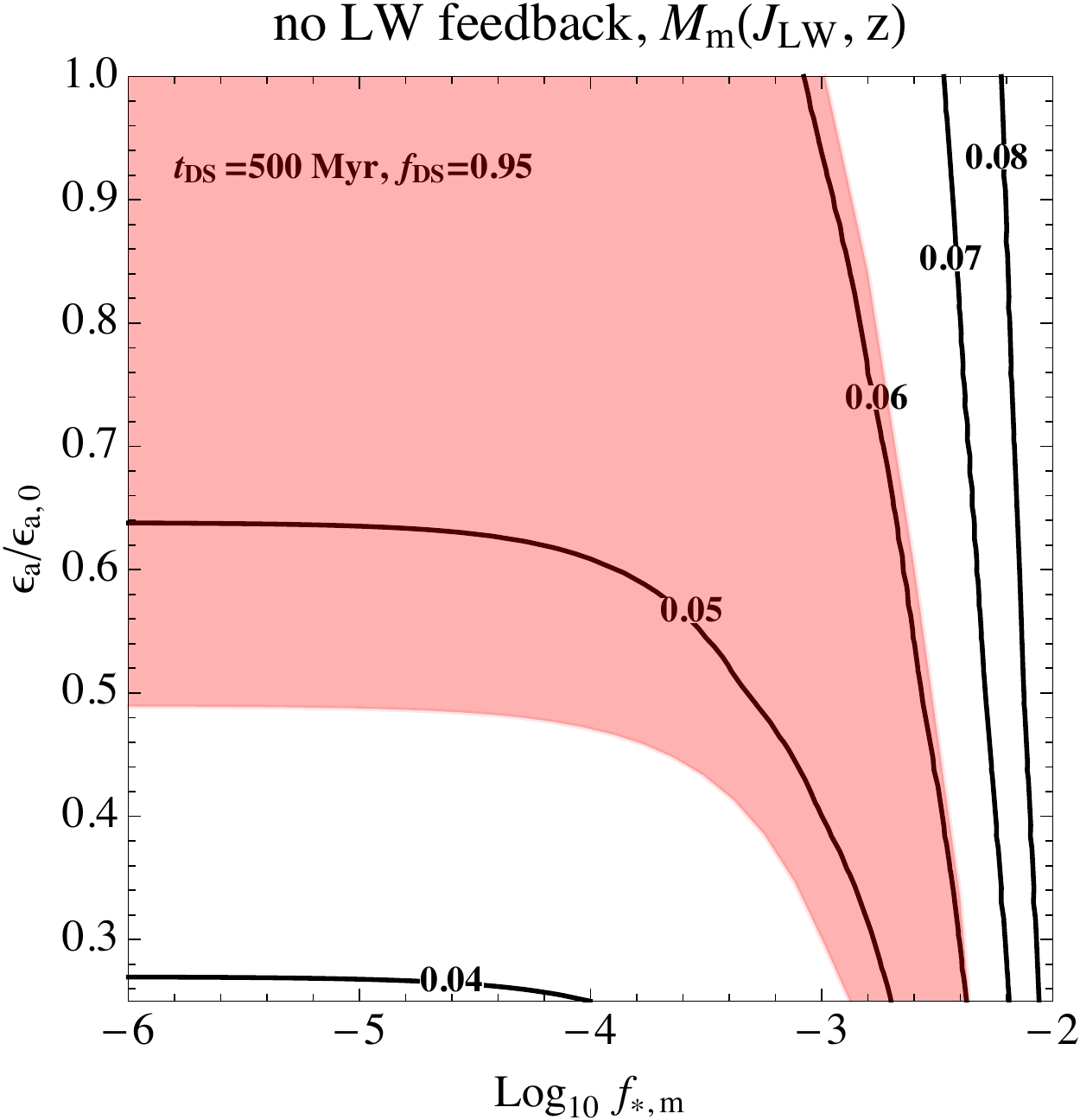}\hspace{3mm}
  \includegraphics[width=0.47\textwidth]{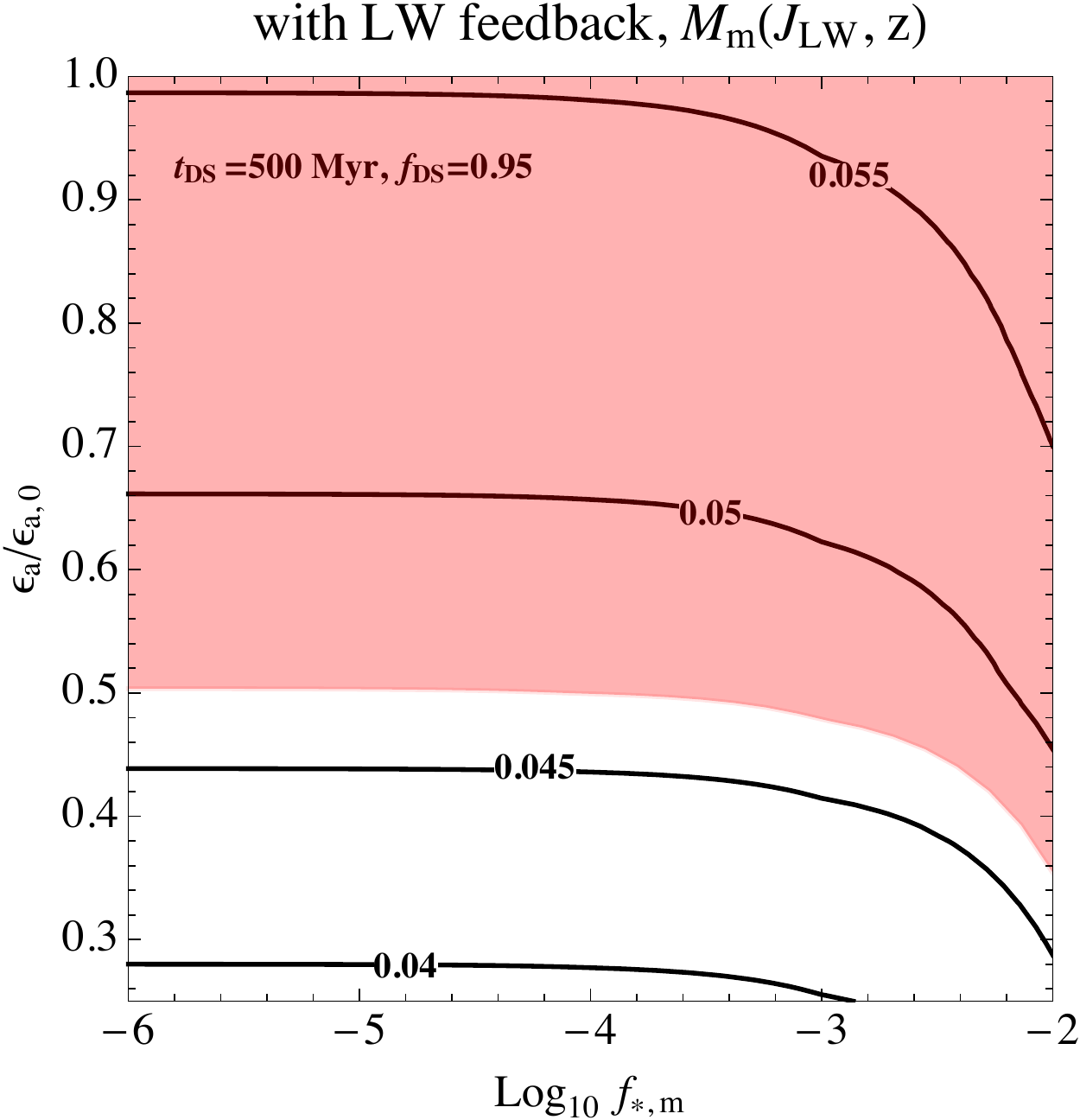}\\\vspace{4mm}
    \includegraphics[width=0.47\textwidth]{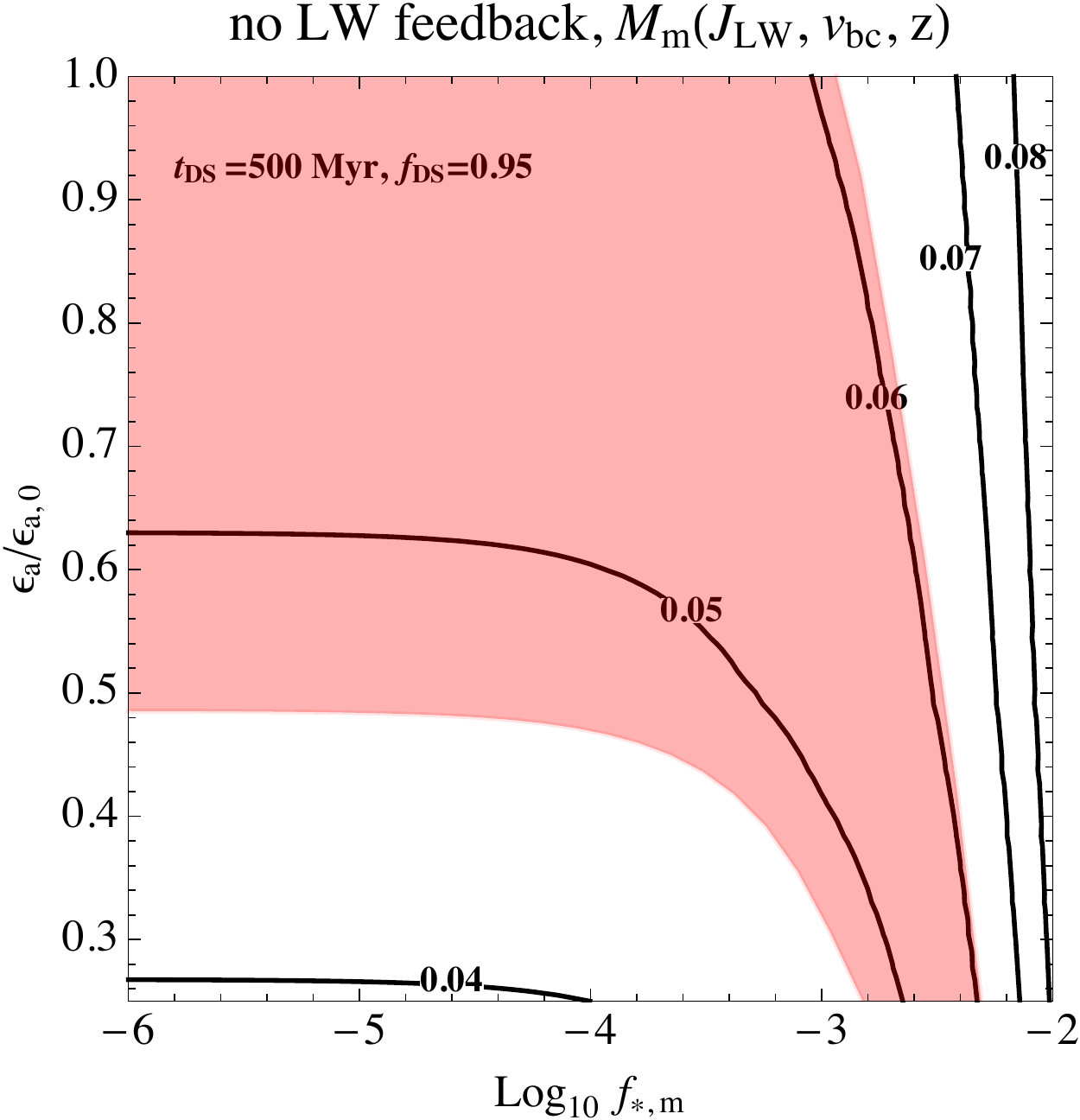}\hspace{3mm}
  \includegraphics[width=0.47\textwidth]{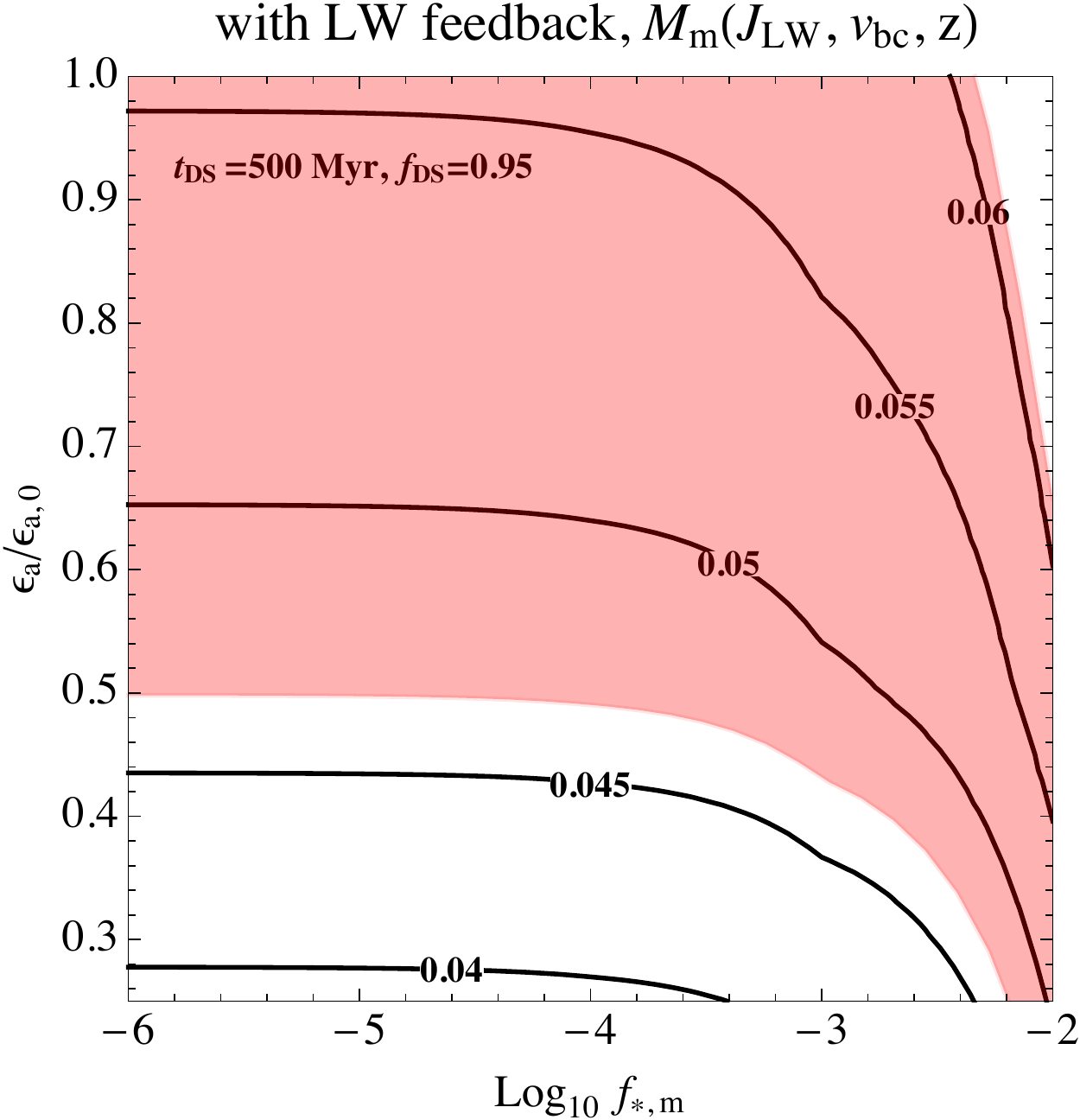}\\
  \caption{contours of the integrated optical depth to
the last scattering surface as a function of $f_{*,\text{m}}$ and $\epsilon_{\text{a}}$ in the presence of DSs with $t_{\text{DS}}=500\,\text{Myr}, f_{\text{DS}}=0.95$. Top (bottom) panel corresponds to the minimum mass of minihalos, $M_{\textrm{m}}$, in which the effect of the baryon-DM streaming velocity, is ignored (included). Left (right) panels show the optical depth when LW feedback is ignored (included). The red shaded regions display $1\sigma$ regions based on the integrated optical depth to the last scattering surface observed by {\it Planck}.}
  \label{fig:tauAstro2}
\end{figure}

In Fig.~\ref{fig:tauAstro2} we present contours of the integrated optical depth to the last scattering surface in the $(f_{*,\text{m}}, \epsilon_{\text{a}}/\epsilon_{{\text{a},0}})$ plane in the presence of DSs with $t_{\text{DS}}=500\,\text{Myr}, f_{\text{DS}}=0.95$. The top (bottom) panel of Fig.~\ref{fig:tauAstro2} corresponds to $ M_\text{m}=M_\text{m}(J_\text{LW},z)$ ($ M_\text{m}=M_\text{m}(J_\text{LW},v_\text{bc},z)$). The left (right) panel of Fig.~\ref{fig:tauAstro2} show contours of the integrated optical depth
to the last scattering surface without (by) including LW feedback.

\begin{figure}[t]
  \centering
    \includegraphics[width=0.55\textwidth]{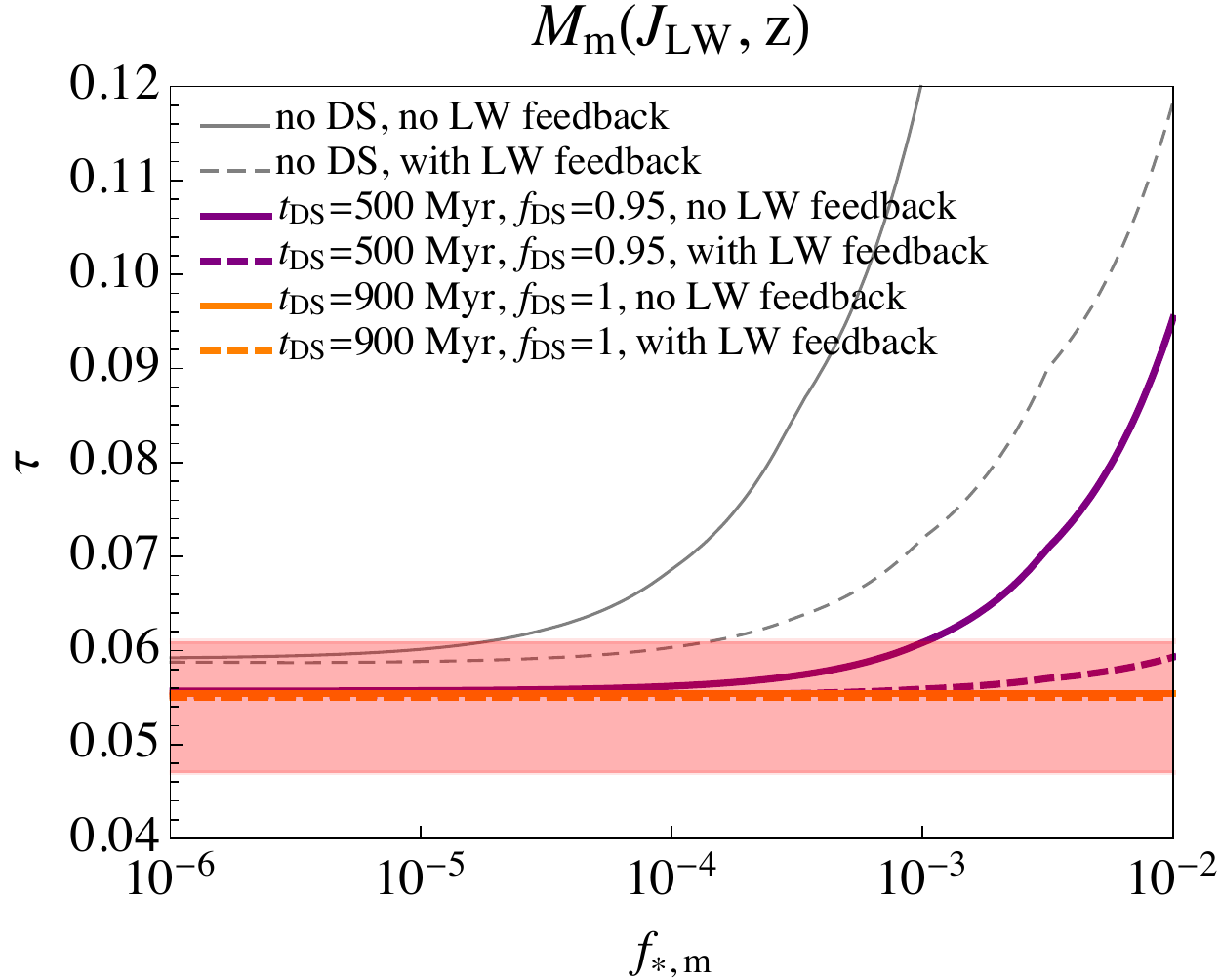}
  \caption{The integrated optical depth to
the last scattering surface as a function of minihalo star formation efficiency, $f_{*,\text{m}}$ when $\epsilon_{\text{a}}=\epsilon_{\text{a},0}$, and $ M_\text{m}=M_\text{m}(J_\text{LW},z)$.
Gray solid (dashed) curve corresponds to a reionization history without DSs and in the absence (presence) of LW feedback, purple solid (dashed) curve represents reionization in the presence of  DSs with $t_{\text{DS}}=500\,\text{Myr}, f_{\text{DS}}=0.95$ without (with) LW feedback, and orange solid (dashed) curve shows the effect of DSs with $t_{\text{DS}}=900\,\text{Myr}, f_{\text{DS}}=1$ on reionization history, without (with) LW feedback. 
The red band displays $1\sigma$ region based on the integrated optical depth to the last scattering surface observed by {\it Planck}.}
  \label{fig:fstarm}
\end{figure}

As Figs.~\ref{fig:tauAstro1} and \ref{fig:tauAstro2} show clearly, while LW feedback by itself can only decrease the integrated optical depth down to $\tau\simeq 0.05$, which demands small values of $f_{*,\text{m}}$ and $\epsilon_{\text{a}}$,
DSs can explain the small value of the measured integrated optical depth for larger values of $f_{*,\text{m}}$ and $\epsilon_{\text{a}}$.
This also is shown in Fig.~\ref{fig:fstarm} which depicts the integrated optical depth to the last scattering surface as a function of $f_{*,\text{m}}$ when $\epsilon_{\text{a}}=\epsilon_{\text{a},0}$ and for $ M_\text{m}=M_\text{m}(J_\text{LW},z)$. 


In Fig.~\ref{fig:fstarm}, the gray solid curve displays the integrated optical depth without DSs and LW feedback; the gray dashed curve shows the effect of adding LW feedback; the purple solid (dashed) curve corresponds to DSs with $t_{\text{DS}}=500\,\text{Myr}, f_{\text{DS}}=0.95$ without (with) LW feedback; and the orange solid (dashed) curve represents DSs with $t_{\text{DS}}=900\,\text{Myr}, f_{\text{DS}}=1$ without (with) LW feedback. Although the impact of the LW feedback can be important for large values of $f_{*,\text{m}}$ (purple solid and dashed curves), it becomes subdominant when increasing the lifetime of DSs or their mass fraction (orange solid and dashed curves).

\section{CONCLUSIONS}
\label{sec:conclusion}
We have studied the effect of DSs on the reionization history of the Universe and the interplay between them and LW feedback in explaining the small value of the integrated optical depth to the last scattering surface measured by {\it Planck}. After modifying a semi-analytical reionization model, which incorporates Pop II stars in atomic cooling halos and Pop III stars in minihalos with LW feedback, to include DSs as the first phase in star formation, we calculated the total ionized filling factor and, subsequently, the CMB optical depth.

To capture the effect of LW feedback on increasing  the minimum mass of minihalos hosting Pop III stars and consequently delaying the formation of these stars, we adopted two representations of the minimum mass of minihalos; the first representation only depends on LW radiation, while the second one depends on LW radiation and the baryon-DM streaming velocity. We showed that these two representations, in the absence of LW feedback, lead to almost the same results for the integrated optical depth in the presence of DSs:  DSs with a lifetime in the range $100\,\textrm{Myr}\lesssim t_{\textrm{DS}}\lesssim1000\,\textrm{Myr}$ and with a mass fraction in the range $0.95\lesssim f_{\textrm{DS}}\lesssim 1$ give rise to an optical depth consistent with {\it Planck} measurements. With the inclusion of LW feedback, the minimum mass of minihalos which ignore the baryon-DM streaming velocity leads to stronger suppression of the integrated optical depth in the presence of DSs than the minimum mass of minihalos which depends on the baryon-DM streaming velocity: DSs  with a lifetime in the range $100\,\textrm{Myr}\lesssim t_{\textrm{DS}}\lesssim1000\,\textrm{Myr}$ and with a mass fraction in the range $0.7\lesssim f_{\textrm{DS}}\lesssim 1$ ($0.9\lesssim f_{\textrm{DS}}\lesssim 1$), generate the optical depth consistent with limits from {\it Planck} for $M_\text{m}(J_\text{LW},z)$ ($M_\text{m}(J_\text{LW},v_\text{bc},z)$).

We also studied the effects of astrophysical parameters including the star formation efficiency in minihalos hosting Pop III stars and the ionizing efficiency of atomic cooling halos hosting Pop II stars. 
We find that for small values of $f_{*,\text{m}}$ the optical depth depends on both $f_{*,\text{m}}$ and $\epsilon_{\text{a}}$, while for larger values of $f_{*,\text{m}}$ the integrated optical depth is almost independent of $\epsilon_{\text{a}}$. 
We showed that in the absence of DSs, LW feedback can reduce the integrated optical depth to be consistent with the observed data by {\it Planck} for $f_{*,\text{m}} \lesssim 10^{-4}$ for the fiducial value $\epsilon_{\text{a}}=\epsilon_{\text{a},0}\,$, and up to $f_{*,\text{m}}\lesssim 2\times10^{-3}$ for $\epsilon_{\text{a}}/\epsilon_{\text{a},0} \sim 0.3$ (see Fig.~\ref{fig:tauAstro1}).

Finally, we demonstrated that the inclusion of DSs can suppress the optical depth further than LW feedback alone, making it possible that the integrated optical depth is consistent with the {\it Planck} data, and can also do so in the absence of LW feedback so long as $f_{\text DS}$ is large enough (see Fig.~\ref{fig:fstarm}).
While LW feedback by itself can maximally decrease the integrated optical depth down to $\tau \sim 0.05$ as discussed above, DSs can easily attain $\tau<0.05$ for a larger range of astrophysical parameters. We note that while the small values of $f_{*,\text{m}}$ required by LW feedback are still consistent with hydrodynamical cosmological simulations~\cite{2020MNRAS.492.4386S}, DSs can accommodate larger/more moderate values of the star formation efficiency, which may become increasingly interesting with future simulations.

The possibility of probing DSs presents opportunities to gain valuable insight about star formation and the nature of non-baryonic DM.
As we demonstrate here, these fascinating objects can address the small value of the integrated optical depth to the last scattering surface, which makes them interesting targets to pursue.  Furthermore, since the optical depth, as a single number, does not uniquely determine the ionization history, additional cosmological observables such as the ground-state hyperfine transition, corresponding to wavelength of 21 cm~\cite{Furlanetto:2006jb,2012RPPh...75h6901P}, can provide a more detailed picture of the role of DSs in the reionization process. There are a variety of ways to search for indirect signals of DS remnants today~\cite{Rindler-Daller:2014uja,Rindler-Daller:2020yqe}.  For DS that survived to lower redshifts, it is even possible that they will be directly observed by the James Webb Space Telescope~\cite{2010ApJ...716.1397F, 2010ApJ...717..257Z,2012MNRAS.422.2164I}.

\acknowledgments
The work of P.G., P.S. and B.S. is supported in part by NSF grant PHY-2014075. The work of E.V. is supported in part by NSF grant AST-2009309. P.G. is very grateful to Prof.\ Masahide Yamaguchi for his generous support under JSPS Grant-in-Aid for Scientific Research Number JP18K18764 at the Tokyo Institute of Technology.  P.S. would like to thank Katherine Freese and Chris Kelso for useful preliminary discussions.

\bibliography{draft}
\end{document}